# Multi-carrier Transport in Epitaxial Multi-layer Graphene

Yu-Ming Lin[1], Christos Dimitrakopoulos[1], Damon B. Farmer[1], Shu-Jen Han[1], Yanqing Wu[1], Wenjuan Zhu[1], D. Kurt Gaskill[2], Joseph L. Tedesco[2], Rachael L. Myers-Ward[2], Charles R. Eddy, Jr.[2], Alfred Grill[1], and Phaedon Avouris[1]

1. IBM T.J. Watson Research Center, Yorktown Heights, New York 10598, USA.
2. Advanced SiC Epitaxial Research Laboratory, U.S. Naval Research Laboratory, Washington, D.C 20375

Variable-field Hall measurements were performed on epitaxial graphene grown on Si-face and C-face SiC. The carrier transport involves essentially a single-type of carrier in few-layer graphene, regardless of SiC face. However, in multi-layer graphene (MLG) grown on C-face SiC, the Hall measurements indicated the existence of several groups of carriers with distinct mobilities. Electrical transport in MLG can be properly described by invoking three independent conduction channels in parallel. Two of these are n- and p-type, while the third involves nearly intrinsic graphene. The carriers in this lightly doped channel have significantly higher mobilities than the other two.

Significant attention has been focused recently on the electrical properties of graphene grown epitaxially on SiC substrates [1,2], because it offers an ideal platform for carbon-based electronics using conventional top-down lithography techniques. Epitaxial graphene is synthesized by thermal desorption of silicon from the top layers of a SiC crystal with the remaining carbon atoms re-bonding to form graphene layers. Recent advances in graphene growth and device fabrication have led to demonstrations of field-effect transistors made from Si-face epitaxial graphene with radio-frequency (RF) performance exceeding that of their Si-based counterparts [3,4].

It has been noted that the layer structure and electrical properties of graphene films grown on Si-terminated (0001) versus C-terminated ($000\bar{1}$) surfaces of the SiC substrate are quite different [5,6]. In particular, graphene growth is much faster on the C-face, so that at the same growth temperature, multiple layers of graphene are more readily formed on the C-face than on the Si-face substrates. Nevertheless, the conductive carbon films produced on either crystal face of SiC possess the same honeycomb lattice structure and the transport characteristics unique to graphene, such as the linear Dirac *E-k* spectrum [7] and the anomalous quantum Hall effect [8]. In addition, there is mounting evidence that multi-layer graphene (MLG) films grown on the C-face SiC are typically rotationally disordered with a band structure very similar to that of an individual graphene monolayer [7,9-10].

The transport properties of graphene are usually studied via Hall effect measurements, which provide information on the carrier mobility and density. Hall measurements performed at a single magnetic field yield a weighted average of carrier mobility and density, and are strictly applicable to homogeneous samples. In this respect, *single-layer graphene* constitutes the ideal system for Hall measurements since it is a truly two-dimensional material with a single conduction (or valence) band at any Fermi energy. As a result, the carrier density and mobility in single-layer graphene have been accurately determined by conventional Hall measurements based on a single-carrier model. However, results obtained from such measurements are not likely to be representative of the transport properties in MLG, which is not necessarily a homogeneous system and where carriers may possess different mobilities in different graphene layers. In the C-face MLG, it was noted that cyclotron resonance measurements typically yield much higher carrier mobilities than those obtained from direct electrical measurements [11, 12], suggesting non-homogeneous carriers in the system. Therefore, to characterize the transport properties of MLG grown on the C-face of SiC, alternative measurements and analysis techniques are required. In this study, we performed variable-field Hall and resistivity measurements on epitaxial graphene, and the results were analyzed with a multi-carrier model. Good agreements were obtained between experimental data and the model, providing further evidence of multi-carrier transport in the C-face grown MLG.

Epitaxial graphene films studied here were synthesized on semi-insulating 4H-SiC (Cree) and 6H-SiC (II-VI, Inc.) substrates. To avoid factors specific to the reactor and/or growth conditions, we investigated graphene samples produced from two laboratories, IBM and NRL, using different synthesis conditions. At IBM, graphene films were grown in an ultrahigh vacuum (UHV) chamber, where the samples were pre-cleaned in disilane at 810°C and grown in Ar at a pressure in the mTorr range. At NRL, a commercial chemical vapor deposition reactor (Aixtron/Epigress VP508) was used to etch ~500 nm of the SiC substrate in $H_2$ at 1600°C and then grow graphene films at 1620°C in a flowing Ar ambient for 120 minutes. The details of the two synthesis protocols are reported elsewhere [13,14]. Metal contacts made of Pd/Au (20/40 nm thick) were patterned using either e-beam or optical lithography and standard lift-off procedures. The Hall bar geometry of the graphene channel was then defined by oxygen plasma etching using PMMA as the etch mask. Figure 1(a) shows the scanning electron microscope (SEM) image of a six-contact graphene Hall bar device fabricated on C-face SiC with a width of 4 μm and length of 30 μm between the voltage probes. Variable-field resistivity and Hall measurements were performed between 0 and 1 Tesla and carried out in high vacuum (~$10^{-8}$ torr) at room temperature using standard lock-in techniques with an excitation current below 1 μA.

Figures 1(b) and 1(c) show the measured sheet resistivity $\rho_{xx}$ and Hall resistance $R_H$ as a function of magnetic field for thin graphene layers grown from Si-face and C-face SiC substrates, respectively. Both graphene samples were produced at IBM. The number of graphene layers in these devices, determined by Raman spectroscopy and optical adsorption of the SiC peak [15], is 1-2 layers on the Si-face and 2-4 layers on the C-face SiC substrates. In both type of devices, the sheet resistivity $\rho_{xx}$ remains nearly constant (within 0.5%) and the Hall resistance, $R_H$, is linearly proportional to the magnetic field, exhibiting excellent agreement with the behavior expected for transport involving only a single type of carrier. In a single-carrier system, these two measured quantities, $\rho_{xx}$ and $R_H$, are determined by the carrier density, $n_H$, and Hall mobility, $\mu_H$, through the relations $R_H = B \cdot (q n_H)^{-1}$ and $\rho_{xx} = (q n_H \mu_H)^{-1}$, where $q$ is the electron charge. Thus, from Figs. 1(b) and (c), the carrier density and mobility can be unambiguously determined by either variable-field or conventional single-field Hall measurement techniques, yielding for the specific samples $n_H = 6.3 \times 10^{12}$ cm$^{-2}$ ($n$-type) and $\mu_H = 800$ cm$^2$/Vs for graphene grown on Si-face SiC, and $n_H = 4.9 \times 10^{12}$ cm$^{-2}$ ($p$-type) and $\mu_H = 2590$ cm$^2$/Vs for C-face graphene sample. Multiple devices have been measured on the same C-face and Si-face substrates, and they all exhibit field-dependent transport behavior well described by the single-carrier model. The relatively high level of doping is commonly observed in graphene layers produced from either Si-face or C-face SiC surfaces [1,5,7,8].

For thicker graphene layers, the results from Hall measurements exhibit drastically different behaviors. Figures 2(a) and (b) show the measured $\rho_{xx}$ and $R_H$ of a Hall bar device fabricated on MLG grown on C-face SiC at IBM. The number of graphene layers on this substrate is estimated to be more than ten based on Raman spectroscopy and optical adsorption of the SiC peak [15]. In Figs. 2(a) and (b), the $\rho_{xx}(B)$ increases with increasing field, varying by more than 15% at $B = 1$ T, and $R_H(B)$ exhibits a non-linear field dependence. The best fitted results using the single-carrier model are shown by the dash-dot curves in Figs. 2(a) and (b), exposing significant deviation from the measured data. This deviation prohibits the unique determination of the carrier density and mobility from conventional Hall measurements performed at a single $B$-field. The non-linear field dependence of $R_H$ indicates the existence of carriers with different mobilities [16,17], and possibly different polarities, in this multi-layer graphene channel.

In an electronic system with N types of distinct mobility carriers, the conductivity tensor σ (the inverse of the resistivity tensor) in a magnetic field applied perpendicular to the transport plane is given by:

$$\sigma_{xx}(B) = \sum_{i}^{N} \frac{n_i q \mu_i}{1 + \mu_i^2 B^2}$$
$$\sigma_{xy}(B) = \sum_{i}^{N} \frac{n_i q \mu_i^2 B}{1 + \mu_i^2 B^2}. \quad (1)$$

For a given $N$, a least-squares fit of Eq. (1) can then be performed on the measured data at various magnetic fields to yield the mobility and density of each carrier type. While an *a priori* assumption on the number of carrier groups $N$ is required for a multi-carrier analysis, in practice, one usually starts with a small test value of $N$ and incrementally increases it until the fit converges to the measured data. For $N = 2$, the multi-carrier analysis yields mobilities of $\mu_1 = 5800$ cm²/Vs and $\mu_2 = 1650$ cm²/Vs with corresponding carrier densities of $n_1 = 1.7 \times 10^{12}$ cm$^{-2}$ (hole) and $n_2 = 7.4 \times 10^{12}$ cm$^{-2}$ (electron). Despite some improvement compared to the single-carrier model, there remains significant discrepancy in terms of the absolute error and the trend in field dependence between the measured data and the calculated $\rho_{xx}$ and $R_H$ (see dashed curves in Figs. 2(a) and (b)), indicating the deficiency of the two-carrier assumption for this sample. However, excellent agreement between the measurements and simulation is achieved by the multi-carrier analysis with $N = 3$, as shown by the solid curves in Figs. 2(a) and (b). The three-carrier model yields carrier mobilities of $\mu_1 = 1840$ cm²/Vs, $\mu_2 = 28500$ cm²/Vs, and $\mu_3 = 6900$ cm²/Vs, with the corresponding carrier densities of $n_1 = 9.9 \times 10^{12}$ cm$^{-2}$ (hole), $n_2 = 5.3 \times 10^{10}$ cm$^{-2}$ (hole), and $n_3 = 3.7 \times 10^{11}$ cm$^{-2}$ (electron). While there is no fundamental limit on the number of carrier groups $N$ in the modeling of MLG, we found that the three-carrier model is able to provide good agreement with experimental results (both $\rho_{xx}$ and $R_H$) from all the C-face MLG samples synthesized at our two laboratories. Another example in Fig. 2(c) shows the Hall measurements of MLG grown on a different C-face SiC substrate. The graphene sample was produced at NRL and consisted of ~ 50 layers of graphene. A similar discrepancy from the single-carrier behavior (shown by dashed curves) is observed, whereas the three-carrier transport model (solid curves) captures all the essential features of the field dependence in the measured data. The extracted carrier mobilities are $\mu_1 = 1500$ cm$^2$/Vs, $\mu_2 = 19300$ cm$^2$/Vs, and $\mu_3 = 12400$ cm$^2$/Vs with corresponding carrier densities of $n_1 = 2.3\times10^{13}$ cm$^{-2}$ (hole), $n_2 = 5.1\times10^{11}$ cm$^{-2}$ (hole), and $n_3 = 7.7\times10^{11}$ cm$^{-2}$ (electron). For comparison, the single-carrier model yields a hole carrier density of $3.5\times10^{13}$ cm$^{-2}$ and a mobility of 1340 cm$^2$/Vs, which clearly does not reflect the carriers present.

The method of multi-carrier analysis using variable-field Hall measurements has been previously employed to characterize transport properties in heterostructures such as AlGaAs [16], InP/InAlGaAs, and bulk InN [17]. In bulk materials, the multi-carrier transport phenomena could arise from the presence of electrons and/or holes in multiple conduction (or valence) carrier pockets in the Brillouin zone, while in a multi-layered structure this could result from carriers with different mobilities residing at separate layers. In MLG grown on the C-face of SiC, the multi-carrier transport is not likely to be attributed to the former, because the simple, conical band structure is largely preserved, as confirmed by photoemission measurements [7]. Instead, the distinct carrier mobilities obtained from the Hall measurements suggest transport involving independent carriers in different layers of the MLG. Although the band structure of the rotationally disordered MLG system is similar to that of an individual monolayer, there is still small coupling between the layers [18], which would allow electrical conduction between them. Furthermore, previous studies have shown that the first few graphene layers close to the SiC substrate are heavily doped, with the doping level decreasing away from the interface [19,20]. Since the carrier mobility of graphene depends on the carrier density [21] and is also sensitive to sources of scattering in the environment, it is expected that graphene layers at different distances from the substrates would possess different carrier mobilities.

Using the multi-carrier analysis, Fig. 3(a) shows the measured Hall mobility as a function of carrier density for 13 graphene devices fabricated from 5 different C-face SiC substrates with more than ten graphene layers, including both samples produced at IBM and NRL. All devices exhibit magneto-transport behaviors similar to those shown in Fig. 2, where good agreement with the data is obtained using the three-carrier model. As shown in Fig. 3(a), the multi-carrier analysis suggests that electrical transport in the measured C-face MLG is dominated by three groups of carriers: one involves a higher density of holes with mobilities below 4000 cm$^2$/Vs, one with a lower density of holes with high carrier mobilities in the excess of 10000 cm$^2$/Vs, and one with electrons with mobilities above 3000 cm$^2$/Vs. We note that while the multi-carrier analysis does not reflect the exact number of graphene layers in the sample participating in transport, the good agreements based on the three-carrier model indicates that the electrical transport of the entire graphene stack can be effectively described by only three conduction channels with different average mobilities and densities. Since epitaxial graphene multi-layers are heavily n-type in the vicinity of the SiC interface [1,19,22] and become nearly intrinsic in subsequent layers [20], we attribute the *n*-type channel (I) to the first few graphene layers near the SiC substrate, the lightly *p*-type channel (II) to the subsequent layers, and the heavily p-type channel (III) to the uppermost layers, as shown in Fig. 3(b). The doping profile of epitaxial MLG is schematically shown in Fig. 3(c). In the lightly-doped channel (II), carriers are effectively screened from the charged scattering sources by the underlying and overlaying graphene layers. This screening preserves the high intrinsic carrier mobility of graphene, allowing for the highest carrier mobility among the three conduction channels. While high carrier mobilities in the quasi-intrinsic part of the MLG have also been previously investigated by far infrared optical absorption techniques [11, 12], the results and analysis shown here present the first evidence of such high mobility carriers in electrical measurements. The origin of the dominant hole carriers (p-type doping) in the uppermost graphene layers is currently unknown, but is likely caused by the adsorbed species on the graphene surface.

In summary, we have performed variable-field Hall and resistivity measurements on epitaxial graphene grown on both the Si- and C-faces of SiC, and shown that transport is essentially a single-carrier type for less than 5 graphene layers, regardless of the SiC face. In thick MLG grown on C-face SiC, however, the Hall measurements suggest that electrical transport involves multiple types of carriers with different mobilities in different graphene layers. It has been argued that the graphene layers near the SiC interface and at the open surface are significantly affected by their environment, leading to high carrier densities and a reduction of the carrier mobilities in those layers. High mobilities are preserved in the quasi-intrinsic graphene region sandwiched between the underlying and overlaying conducting layers.

The authors would like to thank V. Perebeinos, C. Y. Sung, Y. Sun and F. Xia for insightful discussions, M. Freitag for Raman characterization, and B. Ek and J. Bucchignano for the expert technical assistance. J.L. Tedesco is grateful for postdoctoral support from ASEE. This work is supported by DARPA under contract FA8650-08-C-7838 through the CERA program and by the Office of Naval Research.

Figure Captions:

Fig. 1: (a) SEM image of a six-contact Hall bar device fabricated on epitaxial graphene grown on C-face SiC. (b) Measured sheet resistivity $\rho_{xx}$ and Hall resistance $R_H$ as a function of magnetic field for thin graphene layers (1 to 2 ML) grown on Si-face SiC. (c) Measured $\rho_{xx}$ and $R_H$ for thin graphene layers (2 to 4 ML) grown on C-face SiC. Both graphene samples were synthesized at IBM. The solid lines in (b) and (c) represent expected field dependence in a single-carrier model (color on-line).

Fig. 2: Measured (a) sheet resistivity $\rho_{xx}$ and (b) Hall resistance $R_H$ as a function of magnetic field for multi-layer graphene grown on C-face SiC. The experimental data are represented by the symbols. The dash-dot, dashed, and solid curves result from least-squares fitting of the data using the single-, two-, and three-carrier model, respectively. (c) Measured $\rho_{xx}$ and $R_H$ of another MLG Hall bar device fabricated on a different C-face SiC substrate (color on-line). The graphene sample shown in (a) and (b) was produced at IBM while the sample shown in (c) was produced at NRL.

Fig. 3: (a) Hall mobility as a function of carrier density, extracted using the three-carrier analysis, for 13 graphene devices fabricated from 5 different C-face SiC substrates with more than ten graphene layers. (b) Schematics of multi-layer graphene grown on C-face SiC, showing the three dominant channels involving layers with different doping and transport properties. (c) Schematics of the doping profile of epitaxial graphene multilayers (color on-line).

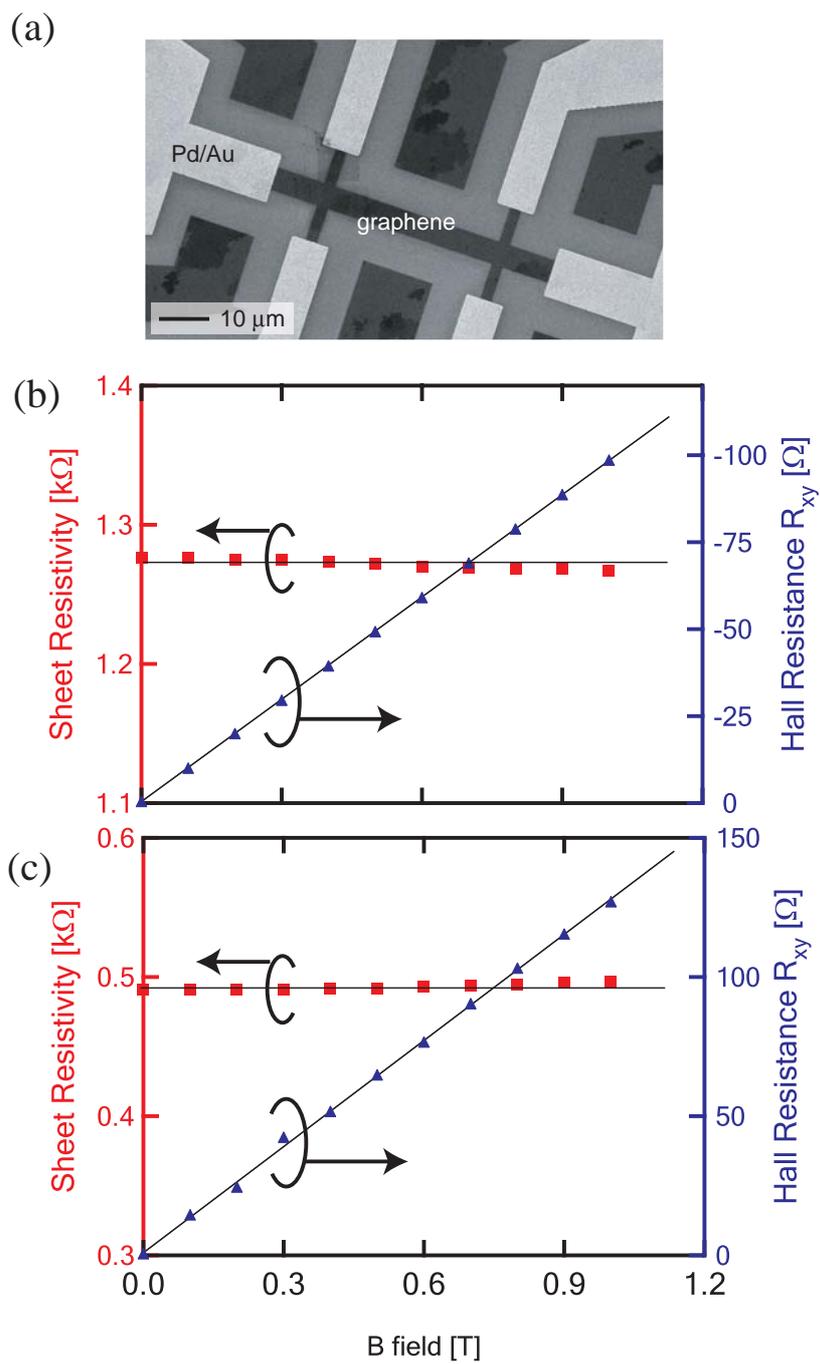

Fig. 1

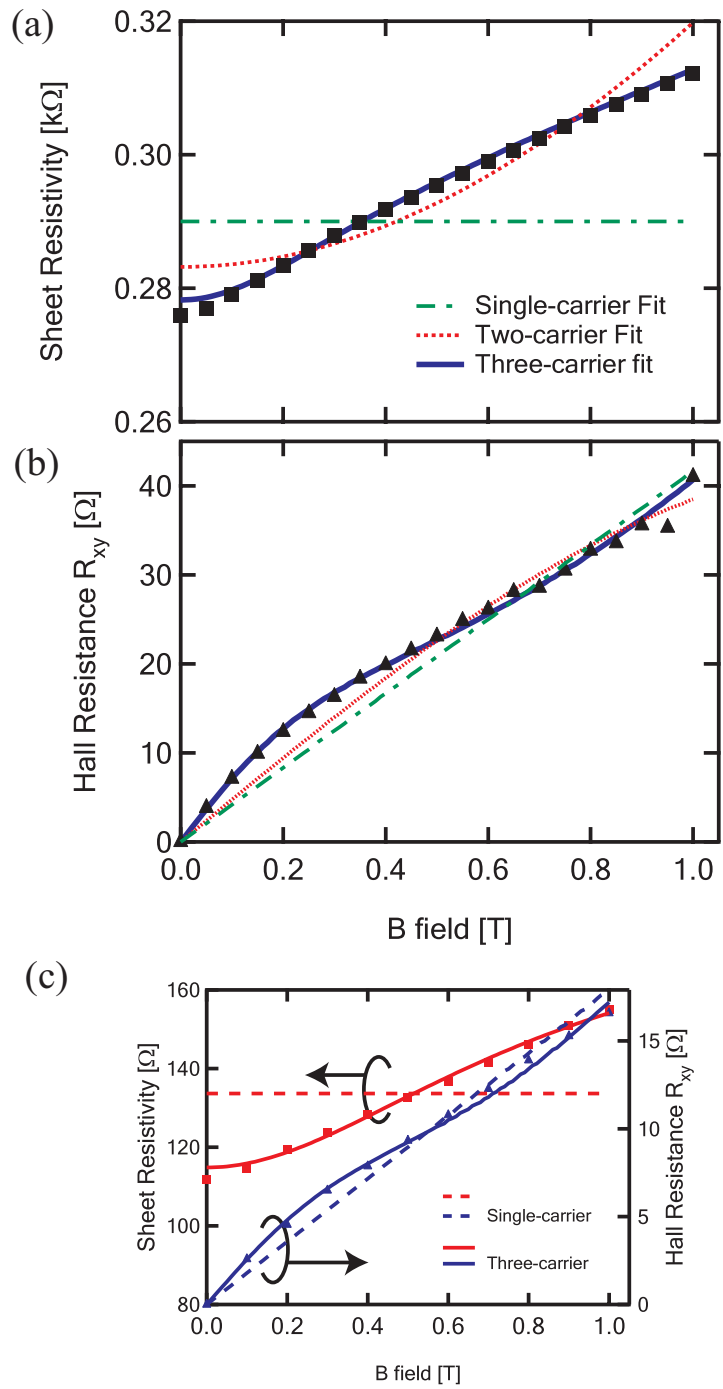

Fig. 2

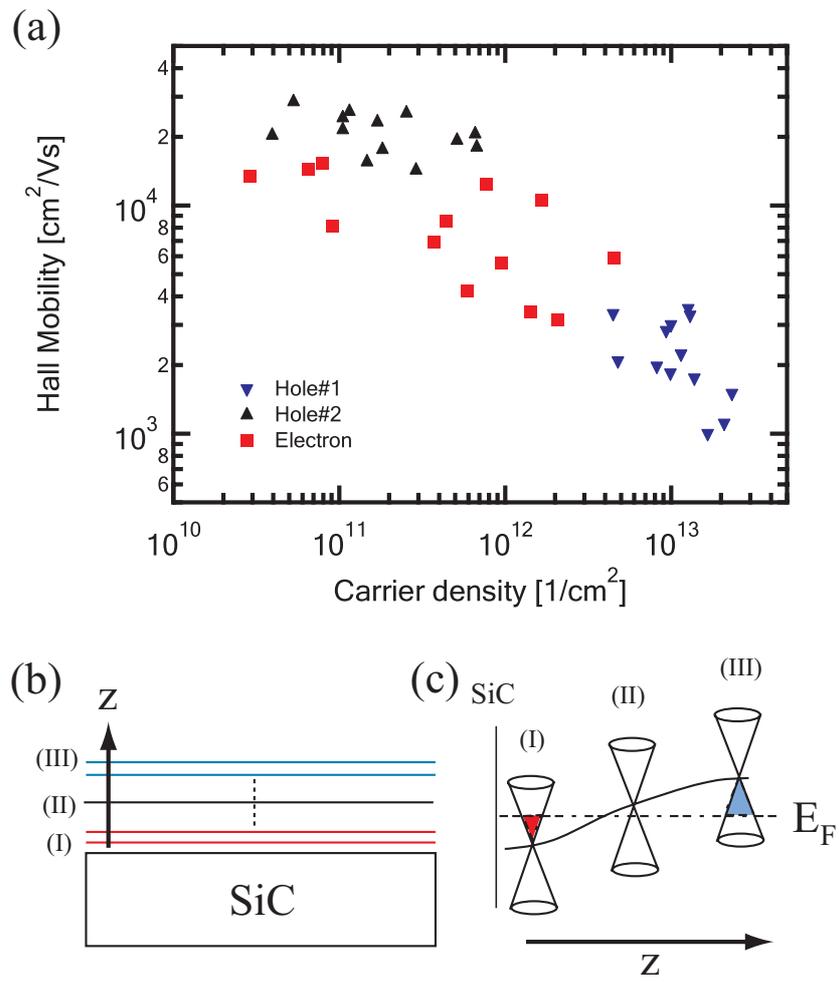

Fig. 3